\def\la{\mathrel{\hbox{\rlap{\hbox{\lower4pt\hbox{$\sim$}}}\hbox{$<$}}}}

\input{aipcheck}

\documentclass[
    ,final            
  ]
  {aipproc}

\layoutstyle{6x9}

\begin{document}

\title{QPOs in CVs: An executive summary}


\author{Brian Warner}{
  address={Dept. of Astronomy, University of Cape Town, Private
Bag X3, Rondebosch 7701, South Africa}
}

\author{Patrick A. Woudt}{
  address={Dept. of Astronomy, University of Cape Town, Private
Bag X3, Rondebosch 7701, South Africa}
}


\begin{abstract}
An overview is given of the properties of the various kinds of quasi-periodic 
luminosity modulations observed in cataclysmic variables (CVs). The two 
principal types, known in the CV literature as dwarf nova oscillations and 
quasi-periodic oscillations, have similarities to the high and low frequency 
quasi-periodic oscillations in X-Ray binaries. There is a further well 
observed class known as longer period dwarf nova oscillations. In CVs the 
observed interrelations between these oscillations suggests a model of 
magnetically controlled accretion onto a rapidly rotating equatorial belt of 
accreted gas. Non-radial oscillations of the central white dwarf are observed 
in some systems.
\end{abstract}

\maketitle


\section{Introduction}

In many respects, the QPOs observed in cataclysmic variables (CVs) parallel 
what are seen in X-Ray binaries (XRB). The two fields have evolved almost 
independently, with slightly different terminologies, but striking 
similarities in properties exist. In order to inform the XRB community 
in a concise way about the CV QPOs we provide a manual of phenomenology, 
together with our current interpretation in terms of magnetic accretion. 
A review of CVs in general is given in \citep{warn95a}; a review of CV QPOs 
is given in \citep{warn04}; here we update the latter with results partly 
taken from our papers \citep{wowa02,wawo02,wawo06,wwp03,pret06,warn08}, and 
including unpublished observations.
   
The story began in 1954 when Merle Walker \citep{walk56} discovered highly stable 71 
s coherent brightness oscillations in DQ Her, the remnant of Nova Herculis 
1934. Despite a search, with the photometric techniques then available he 
could not find any further rapid pulsations in CVs. The advent of pulse 
counting photometry and application of Fourier transforms (FTs) enabled 
low amplitude pulsations to be dragged from the flickering noise inherent 
in the CV mass transfer process \citep{warn72}, showing that 
they are in fact quite common and almost exclusively connected with high 
mass transfer ($\dot{M}$) discs -- i.e. among dwarf novae during outburst 
and among nova-like variables (which are CVs permanently in outburst).
   
At about the same time improved photometric observations of DQ Her showed that 
observed phase and amplitude variations of the 71 s pulsations during eclipse 
\citep{wwhn72} can be modeled as anisotropic radiation from the 
white dwarf primary sweeping around and `illuminating' the concave surface 
of the accretion disc (e.g. \citep{pett80}). A few years later CVs with 
magnetic primaries were discovered, from their optical polarization and X-Ray 
emission, and termed ``polars'' if the field on the primary is strong 
enough ($\sim 10^7 - 10^8$ G) to enforce corotation with the secondary star 
and eliminate the formation of an accretion disc, or ``intermediate polars'' 
if the field is weaker, which allows formation of a disc, truncated at its 
inner edge by the magnetosphere of the primary. It was then realized that 
DQ Her was the first of the intermediate polars, and the anisotropic 
radiation arises in an accretion zone on the primary; the optical radiation 
is reprocessed hard radiation from the accretion zone. Note that 
intermediate polars are the white dwarf equivalents of LMXBs; there are 
no polar equivalents among LMXBs -- their magnetic moments are too small.
About 5\% of isolated white dwarfs show fields over the range 
$10^3 - 10^8$ G \citep{wick00}.

   The QPOs observed in CVs are of at least four kinds, with a range of 
stabilities and periods. We shall describe them in order of discovery.

\section{Dwarf Nova Oscillations}

The first QPOs to be found were those in dwarf novae in outburst 
\citep{warn72} and as a consequence they became known as dwarf 
nova oscillations (DNOs). They have close similarities to the high frequency 
QPOs in XRBs. Their properties can be summarized as follows: 

\begin{figure}
  \includegraphics[height=.3\textheight]{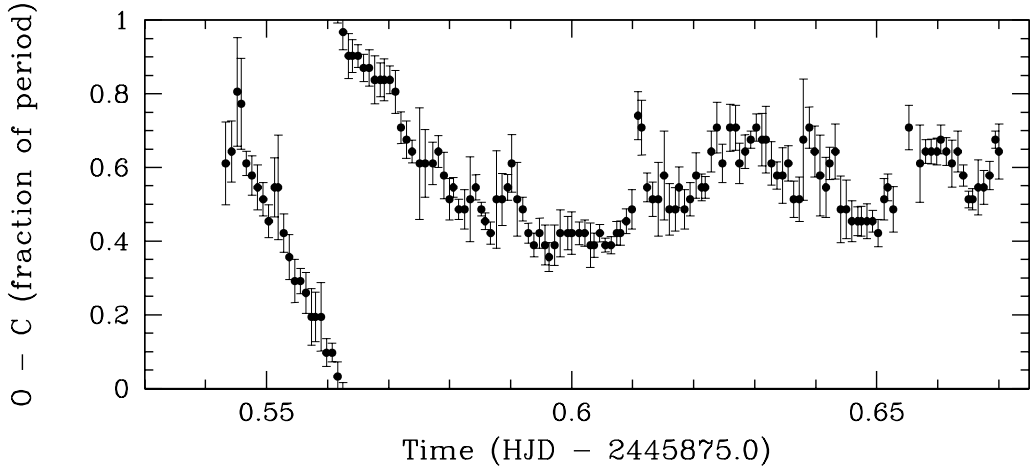}
  \caption{Phase variation of a 26.64-s DNO 
in TY PsA during outburst (run S3362: \citep{wow89}).}
  \label{warnerfig1}
\end{figure}

\begin{figure}
  \includegraphics[height=.5\textheight]{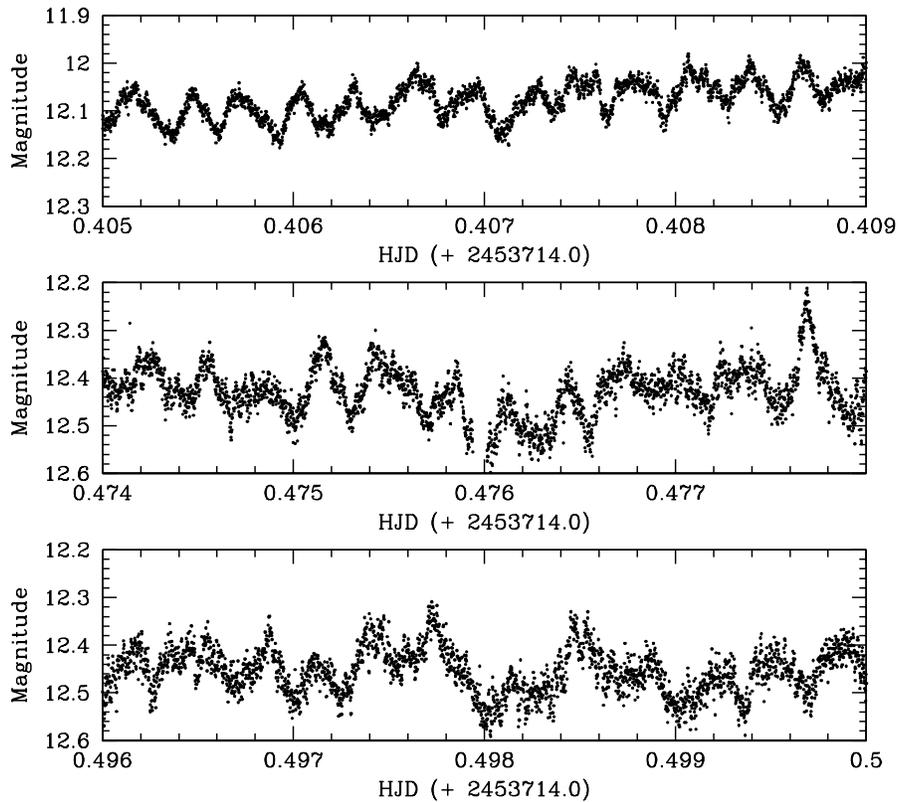}
  \caption{DNOs in VW Hyi during outburst observed with SALT at 80 ms time
resolution.}
  \label{warnerfig2}
\end{figure}

\begin{figure}
  \includegraphics[height=.55\textheight]{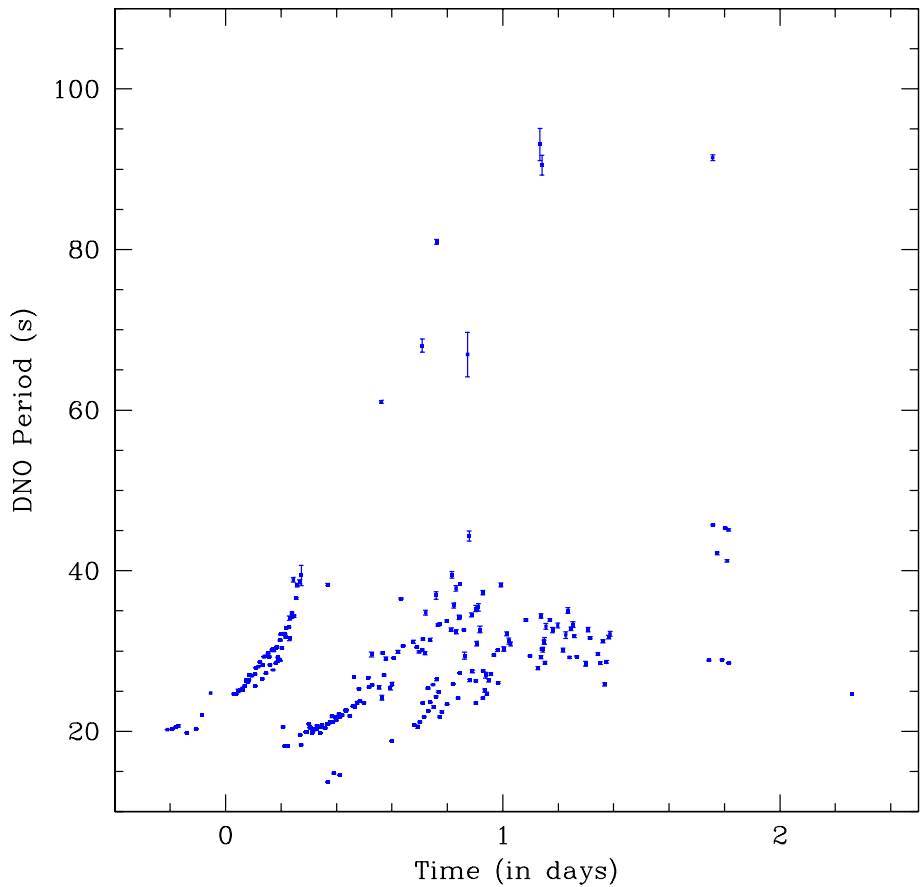}
  \caption{The evolution of the DNO periods at the end of normal
and super outbursts in the dwarf nova VW Hyi. The zero point of the
abscissa is determined from a well-defined averaged template light
curve \citep{wowa02}.}
  \label{warnerfig3}
\end{figure}

\begin{itemize}
\item{They have periods usually in the range 5 -- 40 s, with a characteristic 
range for each CV. For example, the DNO fundamental period in SS Cyg is 
6 -- 11 s.}
\item{There is a period-luminosity relationship, in the sense that minimum 
period corresponds to maximum accretion luminosity, i.e. maximum $\dot{M}$ 
onto the primary. Thus DNO periods decrease on the rising branch of a 
dwarf nova outburst and increase on the descending branch. }
\item{There are sudden small changes in period during the systematic 
variations -- $\sim$ 0.03 s -- but no concomitant detectable luminosity 
changes. Unlike intermediate polars, which have $Q = |\frac{dP}{dt}|^{-1} 
\sim 10^{12}$, DNOs typically have $10^3 < Q < 10^7$. Fig.~\ref{warnerfig1} 
shows an example of an O-C plot of phase as a function of time.}
\item{Although typically of amplitude $<$ 1\% in the optical, and 
detectable only in FTs, occasionally they are directly observable in the 
light curve. Fig.~\ref{warnerfig2} gives an example of DNOs in VW Hyi 
during outburst, as observed with the 11-m SALT reflector. 
The good signal to noise enables the phase of individual pulses to be 
measured (as in Fig.~\ref{warnerfig1}). In FTs of 
lengthy runs of DNOs no first harmonic is detected, but Fig.~\ref{warnerfig2} 
shows that individual pulses may be far from sinusoidal.}
\item{DNOs have been observed in soft X-rays and the EUV in a number of CVs. 
The amplitudes are much larger than in the (reprocessed) optical, occasionally 
reaching 100\%, showing that a large fraction of the accreting gas may pass 
through the modulation machine.}
\item{In SS Cyg at the maximum of some outbursts there is a sudden frequency 
doubling, with the period dropping from 6 s to 3 s. In a few other dwarf 
novae the DNO signal disappears altogether around maximum and reappears later.}
\item{The dwarf nova VW Hyi has been especially well studied (it is 
circumpolar in the southern hemisphere and outburst roughly once a month). 
It is unique in showing large amplitude DNOs during the final decline and 
into the first day of quiescence. DNOs during maximum are very rare; the 
minimum period is 14.1 s, observed both in optical and soft X-Rays. There is 
a sequence of behaviour, repeated every outburst, during which the fundamental 
DNO period increases from $\sim$22 s to $\sim$100 s during $\sim$30 hours, 
but frequency doubling and tripling occur. During the latter parts of this 
evolution the fundamental rarely is present: there are often first and second 
harmonics appearing simultaneously, giving a 3:2 period ratio. 
Fig.~\ref{warnerfig3} shows the evolution of DNOs in VW Hyi.}
\item{During the rapid deceleration phase in VW Hyi its EUV flux is observed 
to drop almost to zero.}
\item{Occasional ``double DNOs'' are observed in the FTs. The DNO acquires 
a companion, at a slightly larger period, which often has a detectable first 
harmonic. Its relationship to the standard DNO is demonstrated in the next 
section.}
\item{The existence of a rotating beam of high energy radiation, as in DQ Her, 
comes from HST spectra of the dwarf nova V2051 Oph \citep{stee01}.}
\end{itemize}

\section{Quasi-periodic Oscillations}

In 1977 it was realized that there are large amplitude oscillations visible 
in some CV light curves, with periods much longer than those of the DNOs, 
which had been overlooked in FTs because they have $Q \sim 5 - 20$ and 
consequently their power is spread over a wide band of frequencies 
\citep{patt77}.  Because of their overt
quasi-periodicity they were designated QPOs; they have some similarities 
to the low frequency QPOs in XRBs.

\begin{figure}
  \includegraphics[height=.2\textheight]{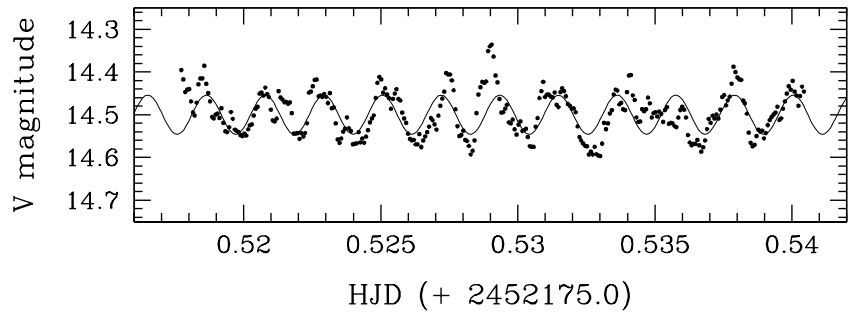}
  \caption{The light curve of WX Hyi showing the 185-s QPO clearly
Superimposed is the result from the non-linear sinusoidal 
least-squares fit (reproduced from \citep{wwp03}).}
  \label{warnerfig4}
\end{figure}

\begin{figure}
  \includegraphics[height=.3\textheight]{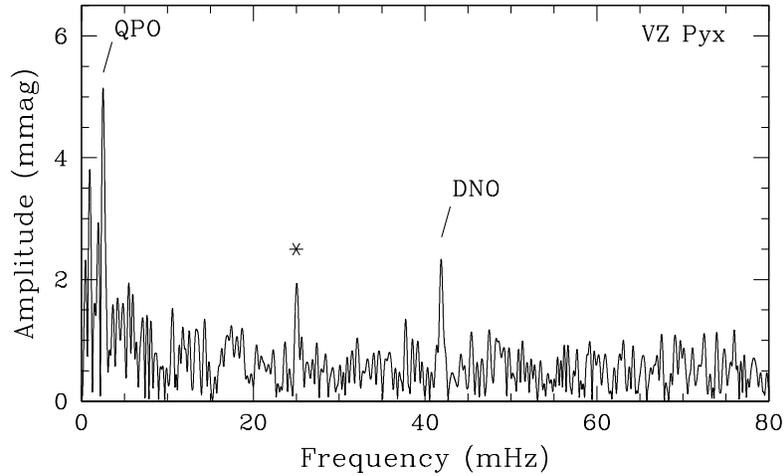}
  \caption{The Fourier transform of run S6569 of VZ Pyx during 
outburst. The QPO and DNO 
periods are marked. The asterisk marks a spurious signal: the telescope's
periodic drive error at 40 s (reproduced from \citep{wwp03}).}
  \label{warnerfig5}
\end{figure}

\begin{itemize}
\item{QPOs also commonly appear in high $\dot{M}$ CVs, but can exist 
independently of the presence of DNOs. Occasionally QPOs are seen in dwarf 
novae during quiescence.  Fig.~\ref{warnerfig4} shows an example of a large 
amplitude QPO that maintains phase coherence for $\sim$10 cycles.}
\item{Harmonics are often seen, including transitions to become entirely 
first harmonic.}
\item{In only one example so far (VW Hyi) has a systematic change of period 
been observed during a dwarf nova outburst. The ratio $P_{\rm QPO}/P_{\rm DNO}$ 
maintains a value $\sim$16 as the DNO period increases rapidly during 
final outburst decline.}
\item{This ratio $P_{\rm QPO}/P_{\rm DNO} \sim 16$ is seen in many other 
high $\dot{M}$ 
systems. We think of this relationship as defining a DNO-related QPO. 
Fig.~\ref{warnerfig5} shows an FT with DNOs and QPOs simultaneously present. }
\item{A significant clue to the nature of the DNO-related QPOs is that 
in several systems the beat period of the double DNOs is equal to the 
observed $P_{\rm QPO}$.}
\item{There is a second kind of QPO that have very long periods 
(typically 1000 -- 2000 s) that are less well studied.}
\end{itemize}

\section{Longer period Dwarf Nova Oscillations}

It was eventually realized that yet another QPO modulation had been 
reported in the literature, and observed by us, but had been previously 
overlooked \citep{wwp03}. As they lie in period between 
the DNOs and the DNO-related QPOs we have called them longer period DNOs 
(lpDNOs).

\begin{itemize}
\item{They are observed only in high $\dot{M}$ CVs, both in dwarf novae 
during outburst and in nova-like variables.}
\item{Occasionally they are directly observable in the light curve.}
\item{Unlike DNOs their periods are only weakly dependent of luminosity 
during outburst. }
\item{It is commonly found that they have periods $\sim$4 times those of 
the DNOs. So a rough relationship is $P_{\rm QPO} \sim 4 P_{\rm lpDNO} 
\sim 16 P_{\rm DNO}$.}
\item{When simultaneously present, DNOs and lpDNOs show independent phase 
and amplitude variations.}
\end{itemize}

\subsection{CV QPOs AND XRB QPOs}

\begin{figure}
  \includegraphics[height=.55\textheight]{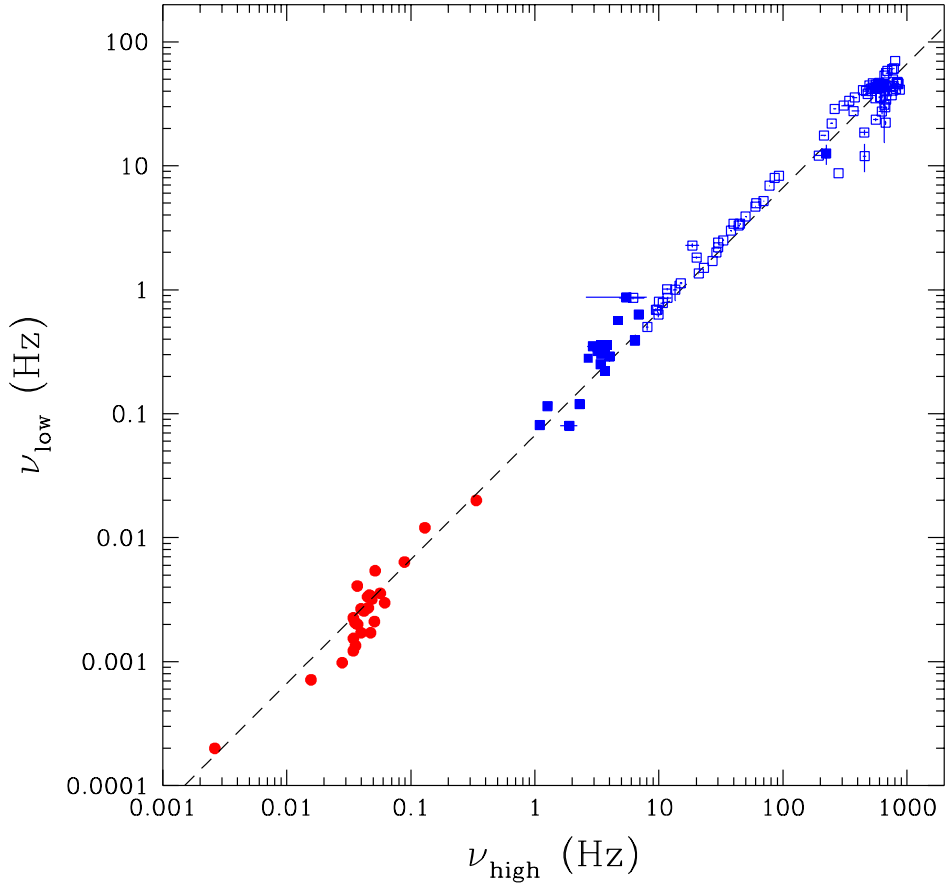}
  \caption{An updated version of the two-QPO diagram for X-ray binaries
(filled squares: black hole binaries; open squares: neutron star binaries)
and 26 CVs (filled circles). Each CV is only shown once in this diagram.
The X-ray data are from \citep{bell02} and were kindly provided
by T.~Belloni. The dashed line marks $P_{\rm QPO}/P_{\rm DNO} = 15$.}
  \label{warnerfig6}
\end{figure}

By considering DNOs to be high frequency QPOs and the DNO-related QPOs to 
be low frequency QPOs we can plot them on the same diagram as  XRB 
observations \citep{wwp03,bell02}; 
an updated version of this graph is shown in Fig.~\ref{warnerfig6}. 
The white dwarf 
relationship is seen to be an extension of that for the more compact 
primaries -- in all cases the ratio of high to low frequency QPOs is 
$\sim$ 15.

\subsection{THE MODEL PROPOSED FOR CV QPOs}

So how do we explain the rich and interconnected phenomenology of CV DNOs? 
The most fundamental issue is the lack of stability of the periods. 
The inverse problem -- the timing stability of white dwarfs like that in 
DQ Her, that have degenerate cores with almost zero viscosity -- was 
discussed by Katz \citep{katz75}, who concluded that a magnetic field $\ge 10^5$ G 
is required to couple the exterior to the central regions of white dwarf, 
resulting in rotation as a solid body, but a dwarf with $B \le 10^5$ G will 
not couple the accretion torque to the interior, allowing a rapidly rotating 
equatorial belt of accreted material to accumulate. It was this concept 
that suggested to Paczynski \citep{pacz78} that the properties of DNOs are related 
to the spin up and spin down of an equatorial belt. This has been elaborated, 
in terms of a low inertia magnetic accretor (LIMA) model (essentially a 
slipping intermediate polar), by Warner \citep{warn95b} and \citep{wawo02}. 
Evidence for hot rapidly spinning equatorial belts on dwarf novae during and 
after outbursts has been deduced from HST spectra 
\citep{sion96,gaens96,cheng97,sko98,sion02}. The mass deposited during a typical 
dwarf nova outburst can be estimated from the total accretion luminosity to be 
$\sim 10^{22}$ g, which has sufficiently low inertia to be tugged around by 
the field coupling it to the inner edge of the disc as the latter moves 
in and out in response to varying $\dot{M}$ during outburst.

The very rapid deceleration of the equatorial belt during the final stages of
a VW Hyi outburst is simply the result of the inner disc radius moving out 
quickly as $\dot{M}$ decreases, leaving a belt rotating so rapidly that the 
co-rotation radius (of the belt) lies beyond the inner edge of the disc 
and so gas is propellered outwards into the outer disc, extracting angular 
momentum from the belt. The EUV flux, which is a proxy for $\dot{M}$ onto 
the white dwarf, falls as a result. Some evidence for propellering in other 
dwarf nova has been found \citep{warn08}.

  We envisage that a field $\sim 10^5$ G will act like a weak intermediate 
polar, as appears to be the case for WZ Sge, which has pulsations at 27.87 s 
(or harmonics) present throughout quiescence and outburst, whereas 
fields $\sim 10^{3-4}$ G may be enhanced by differential shear in the 
equatorial belt during outburst, and fields $\la 10^3$ G will always be too 
weak to channel accretion magnetically and there will be no DNOs observed 
(as is indeed the case for a few well-observed dwarf novae and nova-likes).

  The double DNOs and their relation to the concomitant QPOs suggest an 
analogy to intermediate polars. There, as well as the spin frequency $\omega$ 
of the white dwarf and the orbital frequency $\Omega$, a side band at 
$\omega - \Omega$ is often seen, the result of reprocessing of the white 
dwarf beamed radiation off the secondary (or any other target revolving 
with the orbital period). For the QPOs this required a reprocessing `wall' 
rotating progradely with period $P_{\rm QPO}$, producing QPOs in the light curve 
by alternately partly obscuring and/or reprocessing radiation from the 
central part of the disc, and intersecting the rotating beam. Just such a 
travelling wave is shown by Lubow \& Pringle \citep{lubow93} to be the most likely 
mode to be excited at the inner edge of an accretion disc. There are a 
number of permutations of double DNOs and QPOs possible -- we do not always 
see the QPO in a light curve, even if a double DNO shows that one is present, 
and a QPO wave observed by us is not necessarily intersected by the DNO beam 
to produce double DNOs.

\begin{figure}
  \includegraphics[height=.55\textheight]{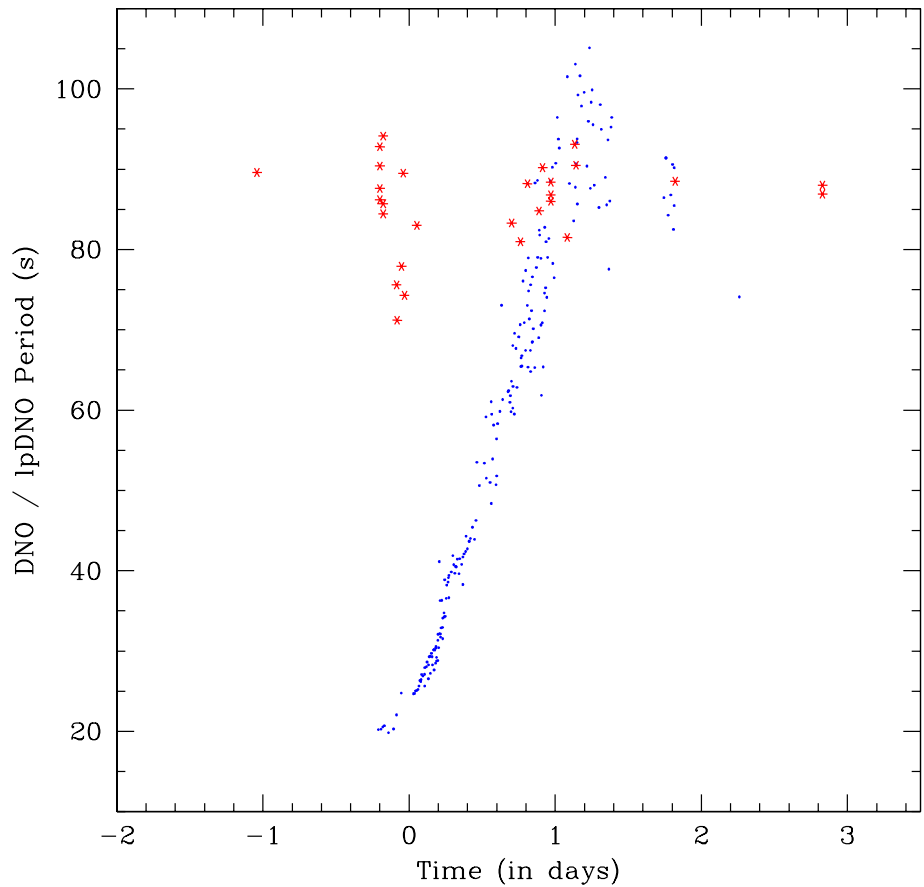}
  \caption{Time-evolution of the observed or implied fundamental
DNO periods (small dots) and lpDNO periods (asterisks).}
  \label{warnerfig7}
\end{figure}

   Finally we come to the lpDNOs -- what is their origin? In many respects 
they behave like standard DNOs, occasionally showing doubling at the QPO 
frequency, but being insensitive to accretion luminosity. A possible clue 
comes from $v \sin i$ measurements made of the primaries from HST 
spectra -- in general the rotation periods of the primaries are twice the 
lpDNO period, consistent with two-pole accretion onto the body of the primary. 
Possibly, therefore, there is accretion from the disc via field lines that 
connect to the primary; alternatively, the source of the lpDNO modulation 
could be in the disc just outside the corotation radius \citep{warn08}. 
The outer parts of the primary are not strongly coupled to the 
interior, so angular momentum of material spreading from the equatorial 
accretion zone will result in differential rotation as a function of latitude, 
perhaps resulting in the variation of lpDNO period. An important clue is seen 
in the overall evolution of DNO and lpDNO periods in VW Hyi 
(Fig.~\ref{warnerfig7}), 
where the fundamental DNO period evolves towards the lpDNO period and ceases 
to increase when the two have converged.

\section{Non-radial oscillations of the primary}

   In addition to the QPOs, which are clearly not situated in the body 
of the primary, multiple oscillations are seen in the primaries of some 
low $\dot{M}$ CVs. Among isolated white dwarfs four instability strips are 
found, within which non-radial oscillations are excited. The first and 
coolest of these was first recognized by \citep{warn72}. 
The surface temperatures of CV primaries are largely determined by 
$\dot{M}$, and it happens that to be in the coolest ($\sim$11\,000 K) 
instability strip requires an $\dot{M}$ that generates an accretion 
luminosity sufficiently low for the white dwarf itself not to be outshone. 
The first of the CVs with pulsating primaries was found in 1997 
\citep{wavZ98}, since when more than a dozen more have been found 
(listed in \citep{marsh06}). Typical pulsation periods lie in the range 
$80 - 1300$ s. We have recently detected what appear to be multiple 
pulsations in two nova remnants, one was a nova in 1986 and the other in 2007. 
If confirmed, these will probably be found to lie in one of the hotter 
instability strips.




\begin{theacknowledgments}
We are grateful to the SALT team for providing the observations on which
Fig.~\ref{warnerfig2} is based; a full description of these will be published
elsewhere.
BW's research is supported by the University of Cape Town. He is grateful to 
the conference organisers for partial financial support that made his 
attendance at the conference possible. PAW's research is supported by the 
National Research Foundation and by the University of Cape Town.
\end{theacknowledgments}

\end{document}